\documentclass{aa} \usepackage{natbib,graphicx}  \usepackage{natbib,graphicx} 

\newcommand{\cog}{\object{Cl~2334+48}}

\begin{document}


\title{Discovery of a cluster of galaxies behind the Milky Way:\\
X-ray and optical observations\thanks{This work is based on the public 
data archive of the XMM-\textit{Newton}, an ESA science
mission with instruments and contributions directly funded by ESA member
states and the USA (NASA).}}

\author{R. Lopes de Oliveira\inst{1,2} \and G.B. Lima Neto\inst{1} \and
C. Mendes de Oliveira\inst{1} \and E. Janot-Pacheco\inst{1} \and C. Motch\inst{2}
}

\offprints{R. Lopes de Oliveira,\\
\email{rlopes@astro.iag.usp.br}}

\institute{
Instituto de Astronomia, Geof\'{\i}sica e Ci\^encias Atmosf\'ericas, 
Universidade de S\~ao Paulo, R. do Mat\~ao 1226, 05508-900 S\~ao Paulo, Brazil 
\and
Observatoire Astronomique, UMR 7550 CNRS, 11 rue de l'Universit\'e, F-67000
Strasbourg, France
}

\date{Received 23 May 2006; accepted 12 July 2006}

\authorrunning{Lopes de Oliveira et al.}
\titlerunning{Discovery of a cluster of galaxies behind the Milky Way}

\abstract{
We report the discovery of \cog, a rich cluster of galaxies in the Zone of
Avoidance, identified in public images from the XMM-\textit{Newton} archive.
We present the main properties of this cluster using the XMM-\textit{Newton}
X-ray data, along with new optical spectroscopic and photometric observations. \cog\
is located at $z = 0.271\pm 0.001$, as derived from the optical spectrum of
the brightest member galaxy. Such redshift agrees with a determination from
the X-ray spectrum ($z$ = 0.263$^{+0.012}_{-0.010}$), in which an intense
emission line is matched to the rest wavelength of the Fe K$\alpha$ complex.
Its intracluster medium has a plasma temperature of $4.92^{+0.50}_{-0.48}$\,keV,
sub-solar abundance of $0.38\pm 0.12\,Z_{\odot}$, and a bolometric
luminosity of $3.2\times 10^{44}$ erg\,s$^{-1}$. A density contrast $\delta =
2500$ is reached in a radius of $0.5\,h_{70}^{-1}\,$Mpc, and the corresponding
enclosed mass is $1.5 \times 10^{14} M_{\odot}$. Optical images show an
enhancement of $g^\prime - i^\prime > 2.5$ galaxies around the central
galaxy, as expected if these were cluster members. The central object is a
luminous E-type galaxy, which is displaced $\sim 40\,h_{70}^{-1}\,$kpc from the
cluster X-ray centre. In addition, it has a neighbouring arc-like feature
($\sim 22\arcsec$ or $90\,h_{70}^{-1}\,$kpc from it), probably due to strong
gravitational lensing. The discovery of Cl~2334+48 emphasises the remarkable
capability of the XMM-\textit{Newton} to reveal new clusters of galaxies in
the Zone of Avoidance.
}

\maketitle

\section{Introduction}\label{sec:intro}

Rich clusters of galaxies have been successfully used as tracers of large-scale structure formation and evolution, which has allowed setting constraints on
various cosmological parameters \citep[see,
e.g.,][]{Bardeen86,Henry00,Rosati02}. In the past, the majority of rich
clusters were first identified in the optical and later observed in X-rays.
However, with the advent of large and deep X-ray surveys, X-ray observations
have become one of the most useful techniques for discovering clusters of
galaxies, especially for intermediate and high-redshift systems
\citep[e.g.][]{Gioia94,Rosati98,Adami00}. On the other hand, X-rays 
may also be useful for detecting clusters near the Galactic plane, where the
increasing number of stars and extinction makes the optical
identification of background galaxies difficult \citep[see, e.g.,][and references
therein]{Kraan00}.

\begin{figure*}
\centering
\includegraphics[width=17.5cm]{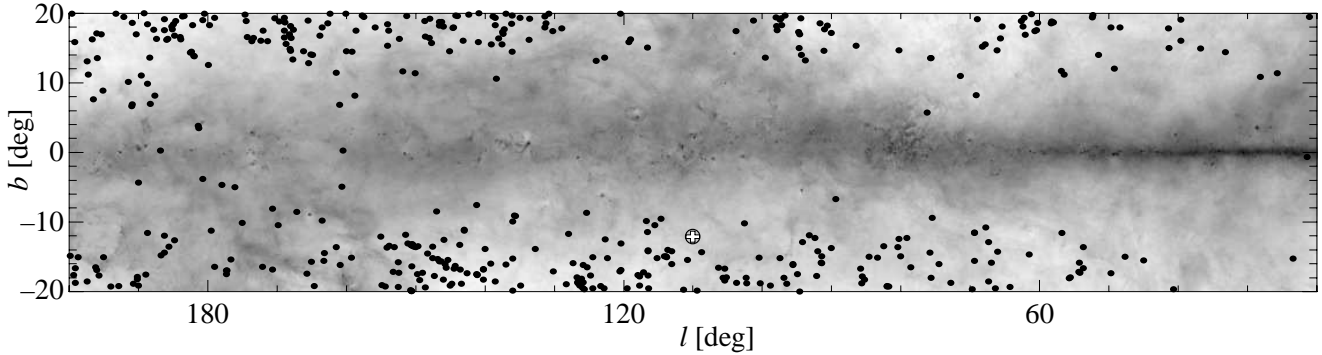}
   \caption[]{Dust map [$E(B-V)$ map from \citet{Schlegel98} in logarithmic
   grey scale] superposed with all known clusters and groups found in NED with
   $|b| < 20^{\circ}$ and $20^{\circ} < l < 200^{\circ}$. \cog\ is
   emphasised with a larger symbol near $l = 110^{\circ}$,~$b = -12^{\circ}$.
   \label{fig:ZoA}}
\end{figure*}

The mapping of the large-scale structures hidden by the Galaxy has profound
cosmological implications, as can be exemplified by the discovery of
the Great Attractor at $(l, b) \approx (320^{\circ}, 0^{\circ})$
\citep{Lynden88, Kolatt95}. Indeed, the mass distribution derived from the
large-scale distribution of galaxies and from the peculiar velocity field of
nearby galaxies and/or clusters is still a matter of debate \citep[see,
e.g.,][]{Tonry00,Hudson04,Mieske05}. It is still not clear whether or not the
bulk flow of galaxies is due mainly to the Great Attractor, or if there is a large contribution from the Shapley supercluster or other more distant
large-scale structures \citep{Nagayama06,Proust06,Radburn06}.

A systematic search for clusters of galaxies in the Zone of Avoidance, $|b| <
20^{\circ}$, was carried out by \citet{Ebeling02}, where they made use of
the ROSAT All Sky Survey Bright Source Catalog
\citep[RASS-BSC][]{Ebeling96,Voges99} coupled with optical and near infrared
follow-ups. A major difficulty with this approach is the low-energy band of
RASS (0.1--2.4\,keV) in which X-ray is strongly affected by the Galactic
absorption, notably when $N_{\rm H} \ga 10^{21}\,$cm$^{-2}$.

In this work we report the discovery of \cog, a rich cluster of galaxies
with $z$ = 0.271, projected onto the Galactic plane and identified in the
XMM-\textit{Newton} data archive. The source is described in Sect.
\ref{sec:source}. The main properties of this object were determined through
the analysis of XMM-\textit{Newton} X-ray data (described in
Sect.~\ref{sec:obs}), complemented by optical data (Sect.~\ref{sec:optical}).
The X-ray analysis is shown in Sect.~\ref{sec:x-ray_analise}, followed by a
mass determination in Sect.~\ref{sec:massdet}. The optical data analysis is
described in Sect.~\ref{sec:OptObs}. For distances and luminosities, we use a
$\Lambda$CDM cosmology with $\Omega_{M}=0.3$, $\Omega_{\Lambda}=0.7$, and
$H_{0} = 70\,h_{70}$km\,s$^{-1}$Mpc$^{-1}$.

\section{The source \cog}
\label{sec:source}

A search in NED\footnote{Nasa Extragalactic Database, www.ned.org} shows that
there are 812 known clusters and groups (221 with measured redshift) in the
Zone of Avoidance. Figure~\ref{fig:ZoA} shows the clusters and groups in a
strip around $l = 110^{\circ}$, $b = 0^{\circ}$ superposed on the dust map
from \citet{Schlegel98}. The number of known clusters and groups drops to 134
(59 with known redshift) in a region limited by $|b| < 12.5^{\circ}$.

Within about 2 arcmin from the well-known \object{Z And} symbiotic star, an X-ray source, \object{1E~2331.6+4834}, is located at RA $=23^{\rm h}34^{\rm
m}02.7^{\rm s}$, Decl. $=+48^{\circ}51'08''$ (hereafter \cog, following the
usual nomenclature), found on a Galactic plane survey with \textit{Einstein}
IPC \citep{Hertz84}. Even after optical and radio follow-ups by \citet{Hertz88}
and \citet{Nelson88}, respectively, the nature of this source remained
unknown. This position corresponds in Galactic coordinates to $l=110.0507^{\circ}$, $b=-12.0743^{\circ}$, where the Galactic extinction is
$E(B-V) = 0.211\,$mag \citep{Schlegel98}.

This source was also observed by the ROSAT PSPC and catalogued as
\object{1WGA~J2334.0+4851}, with an estimated flux of $2.57 \times 10^{-13}$
erg\,cm$^{-2}$\,s$^{-1}$ \citep{White00}. XMM-\textit{Newton} also observed
\object{1E~2331.6+4834} serendipitously in 2001. The MOS and pn sensitivities
and spatial resolution allowed us to identify the source
(\object{XMMU~J233402.7+485108}) as a rich cluster of galaxies. This is thus the
second cluster discovered this way by XMM-\textit{Newton} in the
Zone of Avoidance; the other one was \object{XMMU~J183225.4-103645}, a hot ($kT
= 5.8$\,keV), $z = 0.124$, rich cluster \citep{Nevalainen01}.

\section{Observations and data analysis}

\subsection{XMM X-ray data}\label{sec:obs}

\cog\ was observed twice by XMM-\textit{Newton}, in revolutions 209 on 28
January 2001, and 276 on 11 June 2001. In both observations the symbiotic
system \object{Z And} was the main target. About 80\% of the $\sim 14$\,ks of
the second observation (obsID 0093552801) was lost due to high soft-proton background. We
report only on results obtained from the first EPIC observation (exposure time
of about 24\,ks; obsID 0093552701), which was partially contaminated by solar
particles. This observation was made in \textit{prime full window} mode
with the \textit{medium filter}.

The data were reduced with the Science Analysis System (\textsc{sas})
software v6.5. All EPIC data were reprocessed using the \textit{epproc} and
\textit{emproc} tasks. The data were cleaned following standard procedure,
keeping only standard event grades (patterns 0--12 for MOS1/2 and 0--4 for the
pn; flag = 0 always). Periods of high particle background were filtered out
based on $E > 10$\,keV band light curves. For the MOS detectors, we imposed an
upper threshold of 0.4 count/s and a 1 count/s for the pn. The remaining
usable exposure time was 21.2\,ks for MOS1 and MOS2, and 14.2\,ks for the pn
observations.

Since this cluster is located at a very low Galactic latitude, the use of
standard EPIC background files \citep[e.g.][]{Lumb02} is not recommended.
Therefore we estimated the background using large areas free of cluster emission on the same CCD
chip as the source in each camera, after removing X-ray point sources.

The spectral analysis was performed with the X-ray package \textsc{xspec}
11.3.0. The energy channels were grouped such that each bin contained at
least 25 events. In all cases, the three EPIC spectra were fitted
simultaneously with the same model, but allowing a free normalization
factor between the different instruments.

\subsection{Optical photometry and spectroscopy}
\label{sec:optical}

\cog\ optical imaging was obtained by the XMM-\textit{Newton} Survey
Science Center (XMM-SSC) team in the framework of the XID programme
\citep{Yuan03,Watson01}.
The \object{Z And} field is also part of the XMM-SSC survey of the Galactic plane \citep{Motch03}. 
We report on the public data taken with the optical
mosaic Wide Field Camera (WFC) on the 2.5 m {\it Isaac Newton} Telescope (INT)
with the $g^\prime$ and $i^\prime$ (Sloan system) filters on 16 July 2001. The
$g^\prime$ image has 600\,s exposure and 1\arcsec seeing, while the $i^\prime$
has 1200\,s and a seeing of 1.2\arcsec, both observed in photometric
conditions.

The pre-processed and flat-fielded mosaics have astrometric solutions
that are accurate to better than 0.5", confirmed by us from comparison with the
USNO-A2.0 catalogue \citep{Monet98}. The positions and magnitudes of all
detected objects were derived using the program SExtractor \citep{Bertin96}. Calibration 
to the standard SDSS system was made with the general extinction
coefficients provided by the Wide Field
Survey\footnote{\texttt{http://www.ast.cam.ac.uk/$\sim$wfcsur/technical/photom/
}} (WFS). It was possible to make a star-galaxy separation down 
to $m_{i^\prime}$ = 21\,mag.

A spectrum of the brightest cluster galaxy was obtained in order to check
the redshift derived from the X-ray data. The observation was acquired on 14
May 2005 at the Keck II telescope, equipped with the Echellette Spectrograph
and Imager (ESI) in echelle mode. A 1.25$\arcsec$ slit was used, yielding
a velocity resolution of about 93 km s$^{-1}$. The useful range of wavelength
was $\lambda\lambda$6480--7625\,\AA\ (orders 7 and 8), which corresponds to
$\lambda\lambda$5100--6000\,\AA\ at the galaxy rest
wavelength. A redshift for the
object was obtained from cross-correlation techniques with a stellar spectrum
of \object{HD 19476} (a K0III star), from the Elodie
database\footnote{\texttt{http://atlas.obs-hp.fr/elodie/}}
(Sect.~\ref{sec:OptObs}).

\section{Global properties of the X-ray gas}\label{sec:x-ray_analise}

\subsection{Imaging}\label{sec:imagingX}

\begin{figure}[!tb]
\centering
\includegraphics[width=8.5cm]{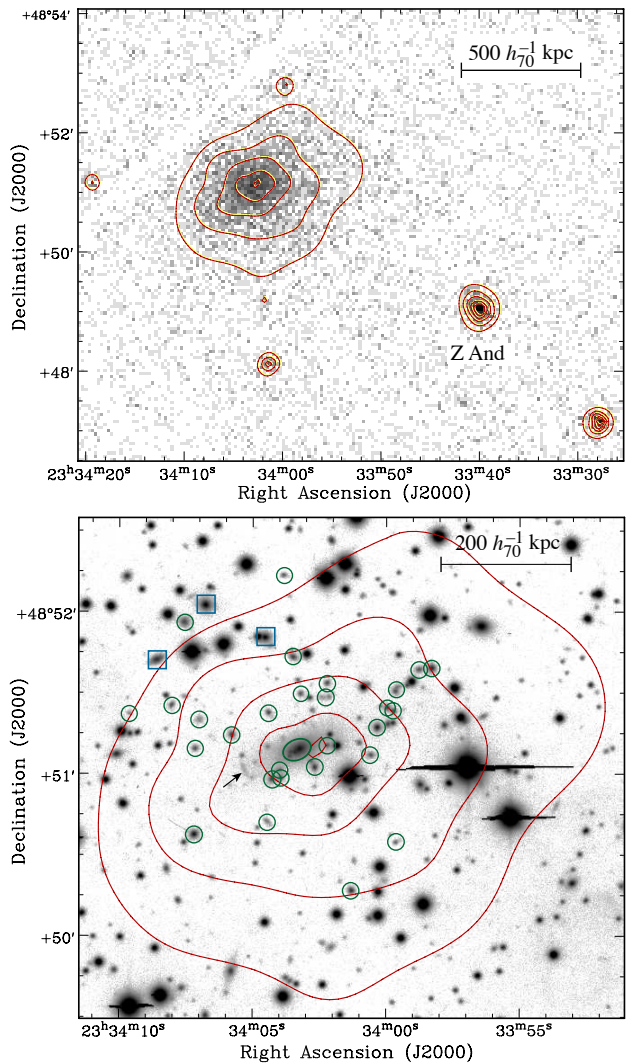}
 \caption[]{\textsf{Top}: EPIC MOS1+MOS2 image of \cog\ in the 0.5--8.0\,keV
 band. The lines are adaptively smoothed contours in logarithmic scales.
 \textsf{Bottom}: The smoothed X-ray contours superposed on the
 $i^\prime$-band image. The circles indicate the identified galaxies inside a
 region of radius $270\,h_{70}^{-1}\,$kpc. Squares mark the next three brightest
 galaxies after the first-ranked galaxy (see Sect.~\ref{sec:OptObs} and
 Fig.~\ref{fig:cmagdiag}). An arc-like feature is indicated with an arrow (see
 Sect.~\ref{sec:OptObs}).
 \label{fig:VibStab}}
\end{figure}

Both EPIC-MOS and pn have a point spread function (PSF) with FWHM $\approx
5''$; however, their PSF have somewhat extended wings and the half energy 
width\footnote{\texttt{http://xmm.vilspa.esa.es/external/xmm\_user\_support/
documentation/uhb/index.html}} is $\approx 14\arcsec$.
The PSF was not taken into account in our analysis.

The X-ray emissivity map shows that \cog\ is quite a regular cluster. 
Figure~\ref{fig:VibStab} displays the X-ray (0.5--8.0\,keV; top panel) and
$i^{\prime}$-band (bottom panel) images, both with the smoothed
X-ray contours overlaid. There are a few features to note. The peak X-ray emission is
displaced from the brightest cluster member by $\sim 10\arcsec$ westwards.
At the cluster redshift ($z = 0.271$; see Sect. \ref{sec:OptObs}), this
offset corresponds to $\sim$ 40\,$h_{70}^{-1}\,$kpc. 
A number of point sources close to \cog\ are identified with active stars and AGNs \citep{Motch06}.
The bottom panel of
Fig.~\ref{fig:VibStab} shows the probable cluster members inside a radius of
270\,$h_{70}^{-1}\,$kpc (see details in Sect.~\ref{sec:OptObs}).

Using the task \textit{ellipse} 
from \textsc{stsdas/iraf}%
\footnote{\textsc{stsdas} is a product of the Space Telescope Science
Institute, which is operated by AURA for NASA. \textsc{iraf} is distributed by
the NOAO, which are operated by the Association of Universities for Research
in Astronomy, Inc., under cooperative agreement with the NSF.}, we derived the
X-ray brightness radial profile $I(R)$. We used a 0.5--8.0\,keV band image,
combining exposure-map corrected MOS1 and MOS2 data, binned such that 1 image
pixel corresponds to $6.4\arcsec$. The CCD gaps and point sources have been
masked out.

\begin{figure}[!tb]
\centering
\includegraphics[width=8.5cm]{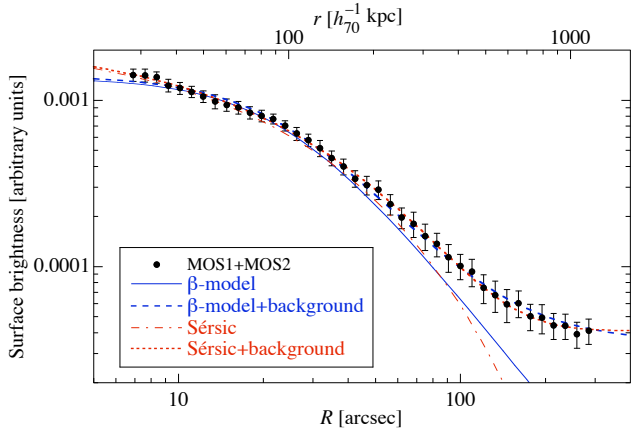}
 \caption[]{X-ray surface brightness fitted with a $\beta$-model and S\'ersic 
 law (see text for details).
 \label{fig:SB}}
\end{figure}

In Fig.~\ref{fig:SB}, we
show the brightness profile, together with two different fits: a S\'ersic
profile \citep{Sersic68} given by
\begin{equation}
    I(R)\,=\,I_{0} \exp\left[ - \left(\frac{R}{a}\right)^{\nu} \right] + B_{1} \, ;
    \label{eq:sersic}
\end{equation}
and a $\beta$-model \citep{Cavaliere76}, in which
\begin{equation}
    I(R)\,=\,I_{0} \left[ 1 + \left(\frac{R}{R_{c}} \right)^{2}\right]^{-3\beta 
    +1/2} + B_{2}\, ,
    \label{eq:beta}
\end{equation}
where $I_{0}$ is the central surface brightness, $R_{c}$ and $a$ are
scale parameters, and $B_{1}$ and $B_{2}$ are constant terms to account for
the background contribution. Table~\ref{tbl:fitSB} summarises the resulting
parameters.
 
\begin{table}[!tb]
 \centering
\caption[]{Surface brightness fitting results, where $R_{c}$ ($\beta$-model) and $a$
(S\'ersic profile) are in $h_{70}^{-1}\,$kpc, and $n_{0}$ is in
$10^{-2}$\,cm$^{-2}$. \label{tbl:fitSB}}
 \begin{tabular}{lcccc}
   \hline
   \hline
    Model        & $R_{c}$ or $a$ & $\beta$ or $\nu$  &  $n_{0}$  \\
   \hline
   $\beta$-model & $98.1\pm 8.5$ & $0.520\pm 0.022$ & $1.49 \pm 0.10$ \\
    S\'ersic     & $65 \pm 14$   & $0.716\pm 0.078$ & $1.30 \pm 0.12$ \\
   \hline
 \end{tabular}
\end{table}

The S\'ersic profile gives a slightly better fit due to the first three
surface brightness data points (the ones most strongly affected by the PSF), 
while the $\beta$-model flattens too much
towards the centre in comparison to the peaked X-ray profile (see
Fig.~\ref{fig:SB}). Notice that the background starts to have an important
effect already at $\sim 30\arcsec$ from the centre. 

In order to estimate the central electronic density, $n_{0}$, which is related
to $I_{0}$, we integrated the bremsstrahlung emissivity along the line-of-sight
within $20\arcsec$ in the central region. The result was compared to the flux
obtained by spectral fitting of the same region (the normalization parameter
of the thermal spectral model in \textsc{xspec}, which is proportional to
$n_{e}^{2}$; Table~\ref{tbl:fitSB}).

\subsection{Spectral analysis}\label{spectralanalysis}

\begin{figure}
\centering
 \includegraphics[angle=-90,width=8.5cm]{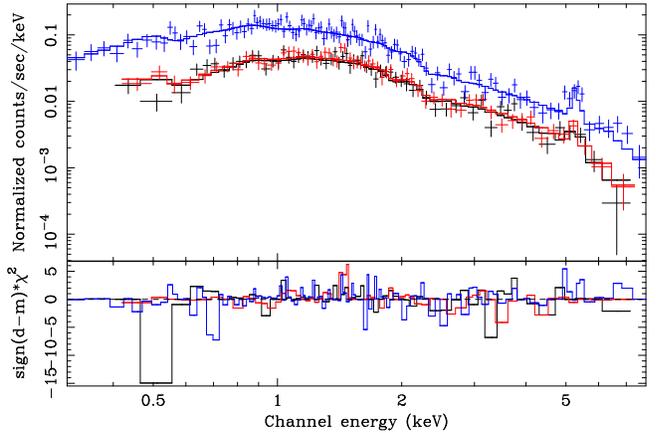}
   \caption[]{EPIC spectra with the best-fit model (solid lines; see Table~\ref{tbl:fitallcluster}).
   \label{fig:fitallcluster}}
\end{figure}

We extracted integrated EPIC spectra in the 0.3--8.0\,keV band from a
circular region of $65\arcsec$ radius ($270\,h_{70}^{-1}\,$kpc at $z = 0.27$),
centred on the cluster X-ray image (see Fig.~\ref{fig:fitallcluster}).
The source's photons correspond to about 90\% of the total good events ($\sim$
1700 photons in each MOS camera, and $\sim$ 3400 in the pn camera), the
remaining being due to background contributions.

The X-ray spectral properties of the ICM can be described very well by the
\textsc{mekal} model \citep[thermal plasma emission;][]{Kaastra93,Liedahl95},
associated to the \textsc{phabs} model \citep{Balucinska92} to account for the
photoelectric absorption (Table \ref{tbl:fitallcluster}). 
We opted for the standard \citet{Anders89} abundance table.

\begin{table*}[!tb]
  \centering
  \caption[]{Spectral fit results with a \textsc{phabs}$*$\textsc{mekal}
  model (see Sect. \ref{spectralanalysis}).
  Uncertainties are at the 90\% confidence level.    
  \label{tbl:fitallcluster}}
  \begin{tabular}{cccccccc}
    \hline
    \hline\\[-2ex]
    $N_{\rm H}$       &  $kT$    & $Z$           & redshift & $f_{X}$
    [0.5--2.0\,keV]$^\ddagger$       & $f_{X}$ [2.0--10.0\,keV]$^\ddagger$        
    & $L_{X}$ [bolom.]$^\ddagger$   & $\chi^{2}_{\nu}/$d.o.f. \\[+0.3ex]
     ($10^{21}$\,cm$^{-2}$) & (keV)    & ($Z_{\odot}$) &          &
     ($10^{-13}$erg\,cm$^{-2}$\,s$^{-1}$) & ($10^{-13}$erg\,cm$^{-2}$\,s$^{-1}$) & ($10^{44}$erg\,s$^{-1}$) & \\     	      
     \hline\\[-2ex]
    $1.6^{+0.2}_{-0.1}$ & $4.84^{+0.50}_{-0.42}$ & $0.40^{+0.13}_{-0.12}$ & $0.263^{+0.012}_{-0.010}$ & 4.6 & 5.8 & 3.0 & 1.09/246 \\
    $1.39^{\dagger}$      & $5.43^{+0.39}_{-0.38}$ & $0.41^{+0.14}_{-0.13}$ & $0.264^{+0.015}_{-0.010}$ & 4.3 & 6.0 & 3.0 & 1.12/247 \\[1ex]
    \hline\\[-2ex]
    $1.6^{+0.2}_{-0.1}$ & 4.92$^{+0.50}_{-0.48}$ & 0.38$^{+0.12}_{-0.12}$ & $0.271^{\ast}$ & 4.6 & 5.8 & 3.2 & 1.09/247 \\
    $1.39^{\dagger}$ & 5.48$^{+0.40}_{-0.39}$ & 0.39$^{+0.13}_{-0.12}$ & $0.271^{\ast}$ & 4.3 & 6.0 & 3.2 & 1.12/248 \\
     \hline\\[-2ex]
  \end{tabular}
  \begin{flushleft}
    \vspace{-2ex}
    $^\ddagger$ corrected for absorptions;
    $^\dagger$ fixed at the Galactic value \citep{Dickey90};
    $^\ast$ fixed at the optical redshift (Sect.~\ref{sec:OptObs}). \\
  \end{flushleft}
\end{table*}

A relatively strong emission line was detected at $5.27^{+0.08}_{-0.10}$\,keV,
with an equivalent width of $669^{+315}_{-284}$\,eV. Setting all parameters free
during the fit, we found that this line is compatible with the Fe K$\alpha$
complex emitted by a thermal X-ray gas located at $z$ =
0.263$^{+0.012}_{-0.010}$. This fit results in a plasma temperature of $kT =
4.84^{+0.50}_{-0.42}$\,keV and a sub-solar abundance of $Z =
0.40^{+0.13}_{-0.12}\,Z_{\odot}$. The resulting hydrogen column density,
$N_{\rm H} = 1.6^{+0.2}_{-0.1} \times 10^{21}$\,cm$^{-2}$, is only slightly
higher than the mean Galactic value obtained in the \citet{Dickey90} HI survey
($N_{\rm H} \sim 1.39 \times 10^{21}$\,cm$^{-2}$). Therefore, there is no
evidence of any notable local excess absorption. To verify this statement, we fitted the whole cluster by fixing the $N_{\rm H}$ to the Galactic value. The
derived temperature, metal abundance, and redshift are compatible with
those of the first model (see Table \ref{tbl:fitallcluster}).

The redshift derived from the X-ray spectrum of Cl~2334+48 agrees very well
with the $z = 0.271 \pm 0.001$ obtained from the optical spectrum of the
brightest cluster member (Sect. \ref{sec:OptObs}). Since there are lower
uncertainties in the optical redshift determination, this redshift was used in
the X-ray spectral fits. As before, similar results are obtained if the
absorption column is taken as a free parameter, or if freezing it to the Galactic
value. Both results are also statistically indistinguishable from those
obtained freeing the redshift in the fits.

\subsubsection{Temperature profile}
\label{sec:TempProf}

In order to determine the temperature profile, we divide the cluster image in a
central circular region and three concentric rings. Their dimensions are such
that the signal-to-noise ratios are approximately constant in each accumulated
data set, as a compromise between the spatial resolution and the quality of
the spectral fitting.

\begin{figure}[!tb]
\centering
\includegraphics[bb=0.8cm 11cm 20.1cm 24.7cm,clip=true,width=8.5cm]{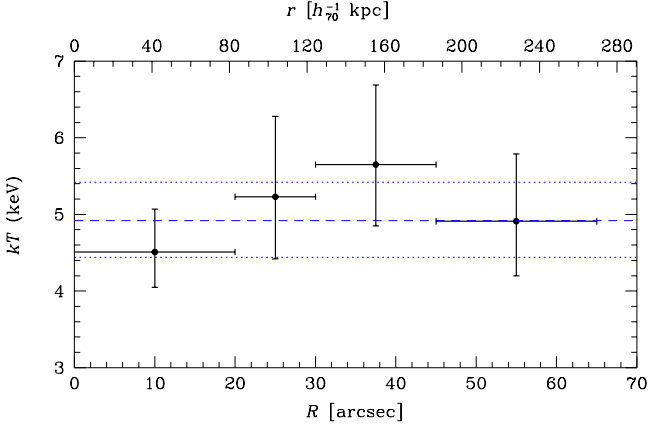}
   \caption[]{Radial temperature profile (see Sect.~\ref{sec:TempProf}). The
   dashed line represents the mean temperature ($kT$ = 4.92\,keV;
   dotted line for their limits) derived from the spectrum extract  within $R <
   65\arcsec$. Error bars are at the 90\% confidence level.   
   \label{fig:tempprofile}}
\end{figure}

The EPIC spectra of each spatial region were fitted with the
\textsc{phabs}$*$\textsc{mekal} model exactly as in
Sect.~\ref{spectralanalysis}, but the absorption column ($N_{\rm H} =
1.6 \times 10^{21}$\,cm$^{-2}$), metallicity ($Z = 0.38\,Z_{\odot}$), and
redshift ($z = 0.271$) were kept fixed to the adopted values of the integrated spectrum
(Table~\ref{tbl:fitallcluster}). The resulting temperature profile is shown in
Fig.~\ref{fig:tempprofile}, where we can see that it shows the
characteristic rise from the centre and then falls outward. However, given the
large error bars and the low spatial resolution, the temperature profile is
still consistent with an isothermal profile at least up to $\sim 70\arcsec$.
Hence, we adopt an isothermal profile with $kT =
4.92^{+0.50}_{-0.48}$\,keV, as derived in Sect.~\ref{spectralanalysis}.

\section{Mass determination}\label{sec:massdet}

\subsection{Gas and dynamical mass}
\label{sec:gdmass}

The gas mass was obtained by simply integrating the gas density obtained in
Sect.~\ref{sec:imagingX}. 
The dynamical mass was computed with the assumption of spherical symmetry and
hydrostatic equilibrium of the X-ray emitting gas, with the density and
temperature profiles obtained above.

We used the
$\beta$-model and S\'ersic law. Both models can be integrated
analytically, resulting in
\begin{equation}
  M(<r)\,=\,4 \pi\, \mu m_{p}\, n_{0} \left\{\!
  \begin{array}{l}
    \displaystyle{\frac{a^{3}}{\nu}\, \gamma\! \left[
    \frac{3-p}{\nu} ,\left(\frac{r}{a}\right)^{\nu} \right]}    \\[2ex]
    \displaystyle{\frac{r^{3}}{3} \, {}_{2}F_{1}\left[ \frac{3}{2}, 
    \frac{3\beta}{2}, \frac{5}{2}, -\left(\frac{r}{r_{c}}\right)^{2} \right]}
  \end{array}
  \right. \! ,
  \label{eq:gasmass}
\end{equation}
where the first line is for the S\'ersic profile and the bottom line for the
$\beta$-model. $\gamma(a,x)$ and ${}_{2}F_{1}(a,b,c,x)$ are the 
incomplete gamma and the hypergeometric functions, respectively \citep[see,
e.g.,][]{Arfken70}.
For both models, $\mu$ is the mean molecular weight ($\approx
0.6$, for a fully ionized primordial plasma) and $m_{p}$ is the proton mass.
For the S\'ersic profile, we also have $2p = 1-0.6097\nu+0.05563\nu^2$
\citep[see][for the deprojection of a S\'ersic profile]{LimaNeto99,Durret05}.

\begin{figure}[!tb]
\centering
\includegraphics[width=8.5cm]{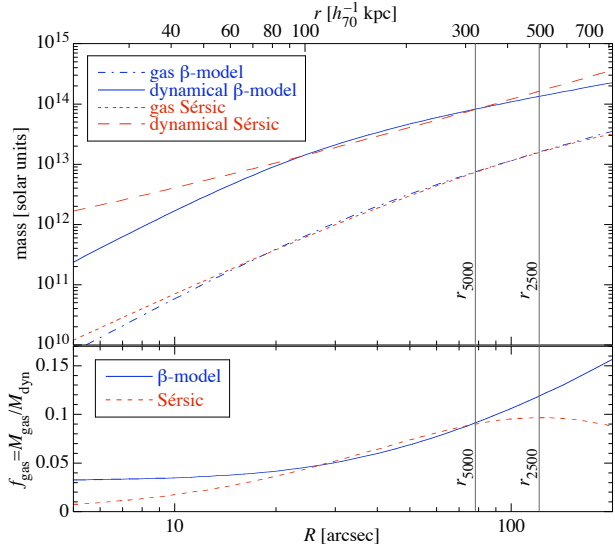}
   \caption[]{\textsf{Top:} Gas and dynamical (total) mass cumulative profiles.
   The vertical lines indicate the $r_{2500}$ and $r_{5000}$ radii (see text). 
   \textsf{Bottom}: the gas mass fraction (baryon fraction minus the 
   contribution from galaxies). 
   \label{fig:mass}}
\end{figure}

Figure~\ref{fig:mass} shows the resulting gas mass profiles, together with the
dynamical mass. Both models predict quite the same gas mass profile in the
range $6\arcsec \la R \la 200\arcsec$ (corresponding to $30 \la r \la
800\,h_{70}^{-1}$\,kpc), where the data are most reliable. Beyond these limits,
the gas mass profiles diverge. The total gas mass at $200\arcsec$ is either
$3.2 \times 10^{13} M_{\odot}$ or $3.5 \times 10^{13} M_{\odot}$ depending on
whether using the S\'ersic profile or the $\beta$-model, respectively. The difference
between the total mass profiles is more marked than the gas profiles,
especially near the centre. At $200\arcsec$, the dynamical mass is either
$3.8\times 10^{14} M_{\odot}$ for the
S\'ersic profile or $2.3 \times 10^{14} M_{\odot}$ for the $\beta$-model.

\subsection{Virial radius and gas mass fraction}\label{sec:dynMass}

\begin{figure}[!tb] 
\centering 
\includegraphics[bb=2.6cm 9.7cm 17.8cm 20.6cm,clip=true,width=8.5cm,width=8.5cm]{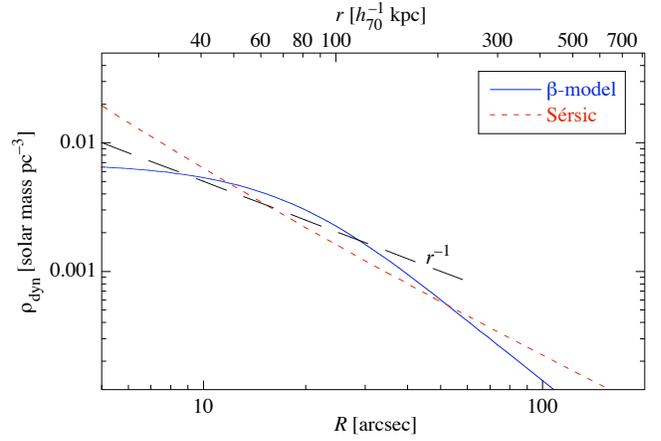} 
  \caption[]{Total mass density profile for both $\beta$-model and 
  S\'ersic law. The long-dashed line shows $\rho$\,$\propto$\,$r^{-1}$ as
  a reference. \label{fig:rhodyn}} 
\end{figure}

We also computed the dynamical mass density as a function of radius, which can,
for example, be compared to the density of dark matter halos formed in
cosmological N-body simulations. In Fig.~\ref{fig:rhodyn} we show the
dynamical mass density compared to a steep $\rho$\,$\propto$\,$r^{-1}$
profile, the same inner slope of the \citet[NFW]{NFW97} ``universal'' profile
for dark halos. 
Note that the inner slope of the total mass density 
is close to $\rho \propto r^{-1}$, like the NFW profile.

While the use of the S\'ersic law resulted in a very steep total density
profile, adopting the $\beta$-model resulted in a central flat total density
profile. However, both profiles are steep up to the X-ray image resolution;
the flattening of the $\beta$-model is actually seen only when the profile is
extrapolated inwards. 

We estimated the radius corresponding to some values of the density contrast $\delta =
\bar{\rho}(r_{\delta})/\rho_{\rm c}(z)$. For $\delta = 200$, we have the usual
$r_{200}$, frequently associated with the virial radius. 
The available data are restricted to about $200\arcsec$, which corresponds to a ratio $\delta \sim
1000$, and therefore well inside the virial radius.

Extrapolating the derived mass profile (Eq.~\ref{eq:gasmass}) from the
available data, we obtained $r_{200} = 3.4\,h_{70}^{-1}\,$Mpc and a corresponding
virial mass of $3.8 \times 10^{15} M_{\odot}$ for the S\'ersic law, or else
$r_{200} = 1.7\,h_{70}^{-1}\,$Mpc and $4.8 \times 10^{14} M_{\odot}$ for the
$\beta$-model. Such a large discrepancy comes from the extrapolation of two
models, owing to the different asymptotical behaviours, and those values are
highly uncertain.

\begin{figure*}
\centering 
\includegraphics[width=17.6cm]{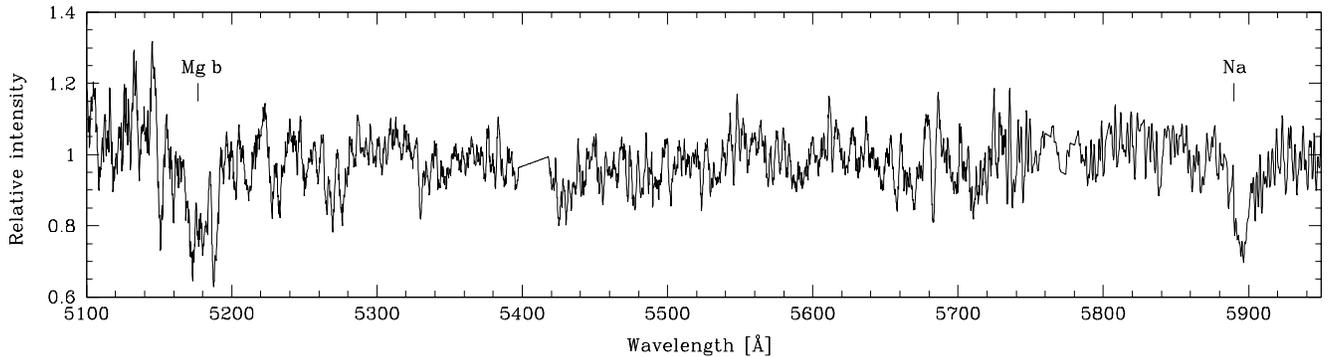} 
  \caption[]{The optical spectrum of the central galaxy, 
   de-redshifted and smoothed with a boxcar filter.
  \label{fig:spectrumgal}} 
\end{figure*} 

Using the mean cluster temperature as an estimator of the virial
radius \citep{Evrard96},
\begin{equation}
    r_{200} = 2.78\,h_{70}^{-1} \left(\frac{kT}{\rm 10\: keV}\right)^{1/2}
    \left(\Omega_{M} (1+z)^{3} + \Omega_{\Lambda} \right)^{-1/2} \, ,
    \label{eq:r200}
\end{equation}
we obtain $r_{200} = 1.7\,h_{70}^{-1}\,$Mpc. The expected mass inside $r_{200}$,
which we take as the virial mass, using the relation $M_{200} \equiv (4\pi/3)
200 \rho_{\rm c} r_{200}^{3}$ is $5 \times 10^{14}\,M_{\odot}$. Both values are
close to that of the $\beta$-model.

For $\delta = 5000$, the equivalent radius can be calculated more precisely.
We obtain $r_{5000} \sim 320\,h_{70}^{-1}\,$kpc for both models, which corresponds to an enclosed mass of $M_{5000} = 8.1 \times 10^{13} M_{\odot}$. For $\delta = 2500$, the radius is still within the X-ray data image, and we have $r_{2500} = 0.5\,h_{70}^{-1}\,$Mpc and $M_{2500} = 1.5 \times 10^{14} M_{\odot}$. These values are very close to the ones from a similar cluster, \object{Abell~1068},
which has $r_{2500} = 490\,h_{70}^{-1}\,$kpc, $M_{2500} = 1.47 \times 10^{14}
M_{\odot}$, and $kT = 4.67$\,keV \citep{Arnaud05}.

The gas mass fraction, $f_{\rm gas}$, is computed simply as the ratio between
the gas mass and the total mass at a given radius, and is related
to the baryon fraction as
\begin{equation}
    \frac{\mbox{baryon fraction}}{\mbox{gas fraction}} = 1 + \frac{M_{\rm
    gal}}{M_{\rm gas}} \, .
    \label{eq:gasfrac}
\end{equation}
Here $M_{\rm gal}$ is the baryonic mass in galaxies, which may be
estimated as $M_{\rm gal} \approx 0.16\,h_{70}^{0.5}M_{\rm gas}$
\citep{White93,Fukugita98}. 

Figure~\ref{fig:mass} (bottom panel) shows the obtained gas mass fraction. For
the $\beta$-model, $f_{\rm gas}$ rises up to the outer limit of the X-ray
data, while this fraction rises up to $\sim r_{2500}$
($\sim 500\,h_{70}^{-1}\,$kpc) and then decreases very slowly for the S\'ersic model. This difference
in behaviour comes from the different form of the dynamical mass for each
model. At $200\arcsec$, $f_{\rm gas}$ is $\sim$ 0.16 for the $\beta$-model and
$\sim$ 0.09 for the S\'ersic profile. These values are close to those obtained
by \citet{Allen02}, $f_{\rm gas} = 0.113 \pm 0.013$ at $r_{2500}$.

Within the interval $5\arcsec \la R \la 200\arcsec$, the derived dynamical
properties, dynamical mass, gas fraction, and total mass density profile could
be considered as normal if compared to the established properties of the regular
rich clusters of galaxies.

\section{Optical properties}\label{sec:OptObs}

The redshift of the central galaxy (RA = $23^{\rm h}34^{\rm
m}03.2^{\rm s}$, Decl. = $+48^{\circ}51'09.3''$) was derived from echelle
spectroscopy, as described in Sect.~\ref{sec:optical}. In particular, the
Mg$_{\rm b}$ band and the Na lines are very prominent in the spectrum
(Fig.~\ref{fig:spectrumgal}). We derived a redshift $z = 0.271 \pm 0.001$,
which confirms the value independently obtained from the X-ray analysis.

\begin{figure}[!tb] 
\centering 
\includegraphics[bb=1.2cm 11.5cm 20cm 23cm,clip=true,width=8.5cm]{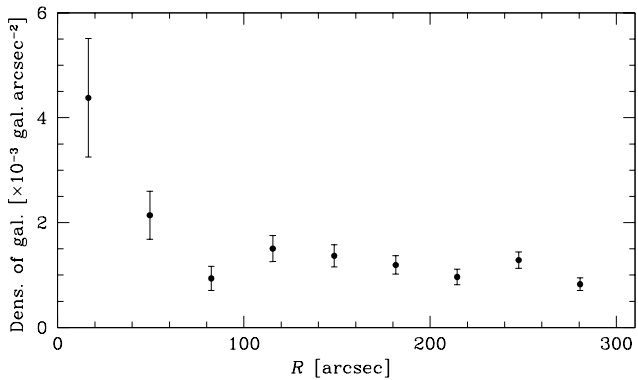} 
  \caption[]{Density of galaxies with 16.3\,$<$\,$m_{i^\prime}$\,$<$\,21 as a 
  function of the distance from the central galaxy in steps of 33$\arcsec$.
  \label{fig:densitygal}} 
\end{figure} 

\begin{figure}[!tb] 
\centering 
\includegraphics[bb=0.9cm 11.5cm 20cm 23cm,clip=true,width=8.5cm]{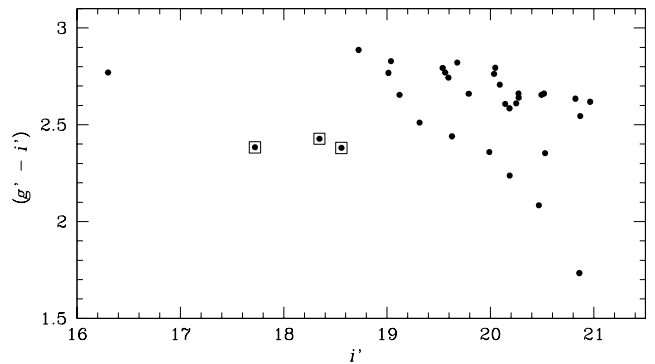} 
  \caption[]{Colour-magnitude diagram in the $g^{\prime}$ and $i^{\prime}$ bands 
  inside a circle of radius 65\arcsec, centred on the brightest cluster galaxy.
  The next three brightest galaxies after the first-ranked galaxy of the cluster 
  are marked with squares (see Sect.~\ref{sec:OptObs}).
  \label{fig:cmagdiag}} 
\end{figure}

The $g^\prime$ and $i^\prime$ images were used to estimate the extent of the
cluster and obtain the colour-magnitude diagram. We show the density of galaxies with observed magnitudes $16.3 < m_{i^\prime} < 21$ in Fig.~\ref{fig:densitygal}, in concentric annuli of increasing radii around
the central galaxy (which has $m_{i^\prime} = 16.3$\,mag). As expected, the
density increases steeply towards the centre of the cluster within a radius of
about $65\arcsec$ (or $270\,h_{70}^{-1}\,$kpc at $z = 0.27$). We then use
all identified galaxies in this region to obtain the colour-magnitude diagram
shown in Fig.~\ref{fig:cmagdiag}. The most probable cluster members have
$g^\prime - i^\prime \ga 2.5$.

The colour-magnitude diagram shows that the next three brightest galaxies
after the first-ranked galaxy of the cluster (Fig.~\ref{fig:VibStab}, lower panel, and Fig.~\ref{fig:cmagdiag}) have
luminosities from four to six times lower than that of the central galaxy.
These three objects are located in the westwards region. Without redshifts, we
can just speculate at this point that these could form a foreground group,
although they do have the right colours to be in the cluster.

\begin{figure} 
\centering 
\includegraphics[bb=1.2cm 11.5cm 20cm 23cm,clip=true,width=8.5cm]{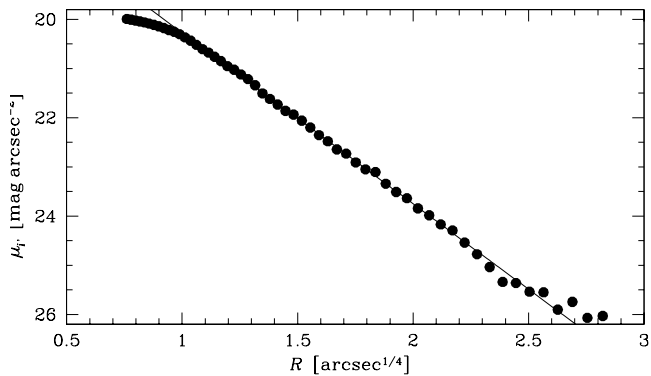} 
  \caption[]{Surface brightness of the central galaxy in the $i^{\prime}$ band.
  The solid line represents a $r^{1/4}$ (de Vaucouleurs) profile. The size of 
  the symbols represents upper limits of the estimated errors of $\mu_{i^{\prime}}$.  
  \label{fig:surfbrigh}} 
\end{figure}

We obtained the surface brightness profile of the central object with only the
$i^\prime$ image (Fig.~\ref{fig:surfbrigh}), since the $g^\prime$ image did not
have sufficient signal-to-noise ratio.  An $r^{1/4}$-profile was fitted to the galaxy surface
brightness profile between 1.5\arcsec and 22\arcsec. There is no significant light excess
over the de Vaucouleurs profile, as one would expect for a cD galaxy.

An arc-like structure is identified in both the $g^\prime$ and $i^\prime$
images (see Fig.~\ref{fig:VibStab}, lower panel). Its relative position,
tangential to the semi-major axis of the expected mass distribution of the
cluster, suggests that this feature could be due to strong gravitational
lensing. Such an arc-like feature is compatible with what is expected for a
distant galaxy at $z$ $>$ 1, lensed by a cluster with mass similar to that
derived for \cog\ in Sect.~\ref{sec:gdmass}.

Assuming that such an arc-like feature is indeed a lensed background galaxy,
we use the singular isothermal sphere (SIS) model to roughly estimate the
cluster mass up to its angular position. We further assume that the arc is
located at the Einstein radius, in this case, $22\arcsec$
($90\,h_{70}^{-1}\,$kpc at the cluster distance) from the centre of the
brightest cluster galaxy. If the background galaxy has $0.8 \la z \la 4$, then
the mass is $(3.5 \la M \la 5.5) \times 10^{13} M_{\odot}$. That is roughly 3
to 4 times the total mass obtained with the X-ray observation (see
Fig.~\ref{fig:mass}).
 
The total mass derived by gravitational lensing tends to be different from
the mass obtained with X-ray observations \cite[e.g.][]{Allen02,Cypriano04}
for hot and/or disturbed clusters. Moreover, the mass derived from lensing
may be higher than the X-ray derived mass due to the mass concentrations along
the line-of-sight of the cluster \citep{Metzler01}. The candidate arc lies
at a position where it can indeed be an image of a lensed background
object. However, only deeper and higher resolution imaging and/or
spectroscopy of the arc itself would confirm its nature.

\section{Summary and conclusions}
\label{discussion}

In this paper we report on the discovery of
Cl~2334+48, a cluster of galaxies in the Zone of Avoidance, from XMM-\textit{Newton}
X-ray images. New photometric and spectroscopic observations of the
system are presented.  The main results of our analysis of the X-ray
and optical data can be summarised as follows.

\begin{itemize}
\item \cog\ is at a redshift of $z = 0.271 \pm 0.001$, based on the optical
spectrum of the brightest cluster member. The redshift derived from the X-ray
spectral fitting is $z = 0.263^{+0.012}_{-0.010}$.

\item The X-ray emission is consistent with an isothermal ICM with $kT =
4.92^{+0.50}_{-0.48}$\,keV, gas metallicity of $0.38\pm0.12$\,$Z_{\odot}$, and
bolometric luminosity of $3.2\times 10^{44}$ erg\,cm$^{-2}$\,s$^{-1}$.

\item There is no evidence of a strong local $N_{\rm H}$ absorption in the ICM.

\item It has, within $800\,h_{70}^{-1}$kpc, a gas mass of $3.2 \times
10^{13}M_{\odot}$ or $3.5 \times 10^{13}M_{\odot}$, a gas fraction of 0.09 or
0.16, and a dynamical mass of $3.8 \times 10^{14}M_{\odot}$ or
$2.3 \times 10^{14}M_{\odot}$ if either a S\'ersic or a $\beta$-model,
respectively, is considered for the gas and total mass cumulative profiles.
The total virial mass of the cluster within the virial radius of $1.7\,h_{70}^{-1}\,$Mpc was determined to be $M_{200} = 5\times 10^{14}M_{\odot}$.

\item There is a significant overdensity of galaxies around the brightest
cluster object, within the inner $65\arcsec$ ($\sim 270\,h_{70}^{-1}\,$kpc).
Most objects are red, with $g^\prime - i^\prime > 2.5$, as expected if these
are members of the cluster.

\item The surface brightness of the central galaxy corresponds to a
$r^{1/4}$ law, indicating that the object is a normal elliptical and not a
cD galaxy.

\item An arc-like structure, probably due to strong lensing, is seen in both 
the $g^\prime$ and $i^\prime$ optical images. This feature is located at $\sim
22\arcsec$ ($90\,h_{70}^{-1}\,$kpc) from the central galaxy of the cluster.
Better images are needed to obtain a secure mass determination of the cluster from the gravitational lensing map.
\end{itemize}

Although Cl 2334+48 was detected in X-ray in the past (see
Sect.~\ref{sec:source}), its nature was unknown. At a Galactic latitude of
$-12^\circ$, an optical identification of the cluster, alone, would be difficult.
Due to the high sensitivity and good spatial resolution of the XMM-\textit{Newton} EPIC cameras, such a
cluster was easily detected, and we were able to derive its main X-ray
properties, as well as its redshift.

The X-ray emission is almost spherically symmetric and the temperature profile
has the usual shape (although the error bars are large) observed in relaxed
clusters (Figures~\ref{fig:VibStab} and \ref{fig:tempprofile}). The X-ray
derived properties based on hydrostatic equilibrium follow the scaling
relations of nearby regular clusters. However, the X-ray peak emission does
not coincide with the brightest member galaxy by an offset of $\sim
10\arcsec$ ($40\,h_{70}^{-1}\,$kpc; see Fig.~\ref{fig:VibStab}). If this
displacement is real, it indicates that this cluster is not quite relaxed and some dynamical event (possibly a sub-cluster merging we may surmise) took
place a few Gyrs ago. The central galaxy is not a cD, owing to the lack of an
extended stellar envelope, which may be interpreted as a sign of the relative
youth of this cluster.

\begin{acknowledgements}

We would like to thank G.J.M. Luna for his participation in this discovery. We
would also like to thank P. Cot\'e and M. West for obtaining the Keck
spectrum of the central galaxy of the cluster, M. Bolte for reducing the spectroscopic data, 
D. Bortoletto for the help with Figures \ref{fig:densitygal},
\ref{fig:cmagdiag}, and \ref{fig:surfbrigh}, and the referee, D. Proust, for his
suggestions and comments. The INT is operated on the island
of La Palma by the Isaac Newton Group in the Spanish Observatorio del Roque de
Los Muchachos of the Instituto de Astrof\'{\i}sica de Canarias. R.L.O.
acknowledges financial support from the Brazilian agencies FAPESP (grant
03/06861-6) and CAPES (grant BEX0784/04-4), and the Observatoire de
Strasbourg. G.B.L.N. acknowledges support from CNPq and CAPES/Cofecub
Brazilian-French collaboration. C.M.d.O. and G.B.L.N. would like to thank support from
FAPESP through the Thematic Project 01/07342-7.

\end{acknowledgements}

\end{document}